\documentclass[structabstract]{aa}  
\usepackage{graphicx}
\usepackage{txfonts}
\usepackage{natbib}
\begin{document}
\title{Fragmentation, infall, and outflow around the showcase massive protostar NGC7538\,IRS1 at 500\,AU resolution\thanks{Based on observations carried out with the IRAM
    Plateau de Bure Interferometer. IRAM is supported by INSU/CNRS
    (France), MPG (Germany) and IGN (Spain). The data are available in
    electronic form at the CDS via anonymous ftp to
    cdsarc.u-strasbg.fr (130.79.128.5) or via
    http://cdsweb.u-strasbg.fr/cgi-bin/qcat?J/A+A/}.}


   \author{H.~Beuther
          \inst{1}
          \and
          H.~Linz
          \inst{1}
          \and
          Th.~Henning
           \inst{1}
           }
   \institute{$^1$ Max-Planck-Institute for Astronomy, K\"onigstuhl 17,
              69117 Heidelberg, Germany, \email{name@mpia.de}}

   \date{Version of \today}

\abstract
{}
{Revealing the fragmentation, infall, and outflow processes in the
  immediate environment around massive young stellar objects is
  crucial for understanding the formation of the most massive stars.}
{With this goal in mind we present the so far
  highest spatial-resolution thermal submm line and continuum
  observations toward the young high-mass protostar NGC7538\,IRS1. Using
  the Plateau de Bure Interferometer in its most extended
  configuration at 843\,$\mu$m wavelength, we achieved a spatial
  resolution of $0.2''\times 0.17''$, corresponding to $\sim$500\,AU
  at a distance of 2.7\,kpc.}
{For the first time, we have observed the fragmentation of the dense inner
  core of this region with at least three subsources within the inner
  3000\,AU.  The outflow exhibits blue- and red-shifted emission on
  both sides of the central source indicating that the current
  orientation has to be close to the line-of-sight, which differs from
  other recent models. We observe rotational signatures in
  northeast-southwest direction; however, even on scales of 500\,AU,
  we do not identify any Keplerian rotation signatures.  This implies that
  during the early evolutionary stages any stable Keplerian inner disk
  has to be very small ($\leq 500$)\,AU).  The high-energy line
  HCN$(4-3)v_2=1$ ($E-u/k=1050$\,K) is detected over an extent of
  approximately 3000\,AU. In addition to this, the detection of
  red-shifted absorption from this line toward the central dust
  continuum peak position allows us to estimate infall rates of $\sim
  1.8\times 10^{-3}$\,M$_{\odot}$yr$^{-1}$ on the smallest spatial
  scales.  Although all that gas will not necessarily be accreted onto
  the central protostar, nevertheless, such inner core infall rates
  are among the best proxies of the actual accretion rates one can
  derive during the early embedded star formation phase.  These data
  are consistent with collapse simulations and the observed high
  multiplicity of massive stars.}
{}
\keywords{Stars: formation -- Stars:
              early-type -- Stars: individual: NGC7538\,IRS1 -- Stars:
              massive}

\titlerunning{Fragmentation, infall and outflow in NGC7538\,IRS1}

\maketitle

\section{Introduction}
\label{intro}

High-mass stars are known to usually form as multiple objects in the
center of dense clusters (e.g., \citealt{zinnecker2007}). With their
large energy output throughout their whole lifetime and their
clustered nature, they are among the most important ingredients of the
Milky Way and extragalactic systems. However, their actual formation
processes are still not well constrained (e.g.,
\citealt{mckee2007,beuther2006b,zinnecker2007}). Particularly
important questions relate to the fragmentation of the innermost dense
cores and the formation of massive accretion disks (e.g.,
\citealt{krumholz2006b,krumholz2009,beuther2007d,bontemps2010,peters2010b,commercon2011,kuiper2011,kuiper2012}).
While Keplerian accretion disks have been found from low- to
intermediate-mass star-forming regions (e.g.,
\citealt{simon2000,schreyer2002,cesaroni2005}), high-mass star-forming
regions have so far usually exhibited non-Keplerian rotation
signatures on scales between a few 1000\,AU and 0.1\,pc (e.g.,
\citealt{cesaroni2007,fallscheer2009,beltran2011}). Since collimated
outflows also exist in high-mass star formation (e.g.,
\citealt{beuther2002a,beuther2002d,zhang2005}), and these outflows
require stable inner accretion disks for the launching (e.g.,
\citealt{vaidya2009}), Keplerian disks are also expected around young
high-mass stars with masses $>$8\,M$_{\odot}$. The fact that the
latter are still observationally not detected in thermal line emission
indicates that these stable Keplerian structures are likely smaller
than the previous spatial resolution limit of thermal line emission on
the order of several 1000\,AU.  Interestingly, our target source
NGC7538\,IRS1 exhibits velocity signatures consistent with a Keplerian
disk observed in non-thermal maser emission on much smaller spatial
scales \citep{pestalozzi2004}. However, these maser features are
controversially discussed and are ambiguous (see next paragraph,
\citealt{debuizer2005,pestalozzi2009}).

\begin{figure}[ht!] 
\includegraphics[width=0.48\textwidth]{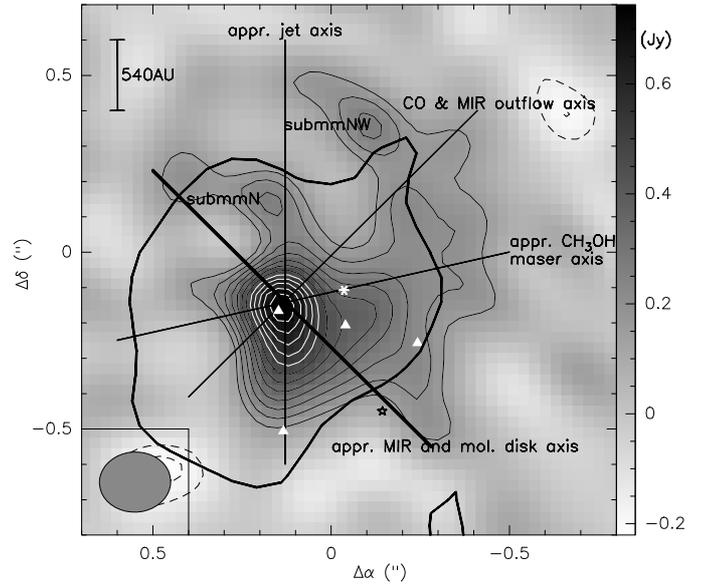}
\caption{PdBI 843\,$\mu$m continuum image toward NGC7538\,IRS1. The
  full line contour levels start at $3\sigma$ values and continue in
  $1\sigma$ steps of 50\,mJy\,beam$^{-1}$ (black contours from 3 to
  9$\sigma$, white contours from 10 to to 14$\sigma$). The dashed
  contour shows the same negative levels. The thick full line presents
  the $4\sigma$ contour ($1\sigma\sim 29$\,mJy\,beam$^{-1}$) from the
  1.36\,mm data presented in \citet{beuther2012c}.  Several potential
  disk and outflow axes reported in the literature are presented
  (\citealt{davis1998,debuizer2005c,sandell2009,pestalozzi2004,pestalozzi2009,sandell2010},
  see Introduction for more details). The thick line presents the axis
  along which the position-velocity diagrams in Figure \ref{pv} are
  conducted. The northern and northwestern peaks submmN and submmNW
  are labeled, and the open star, the six-pointed star, and the white
  triangles mark the positions of the OH, H$_2$CO, and additional
  CH$_3$OH masers \citep{argon2000,hoffman2003,minier2001}. A
  scale-bar and the synthesized beam ($0.2''\times 0.17''$) are shown
  as well. The 0/0 position is the phase reference center in section
  \ref{obs}.}
\label{cont}
\end{figure}

Addressing this spatial resolution challenge, we studied one of the
best northern hemisphere high-mass accretion disk candidates
NGC7538\,IRS1 with the Plateau de Bure Interferometer (PdBI) in the
1.3\,mm wavelength window at $\sim 0.3''$ \citep{beuther2012c}. At the
approximate distance of $\sim 2.7$\,kpc
\citep{moscadelli2009,puga2010}, this corresponds to a linear
resolution of $\sim 800$\,AU.  NGC7538\,IRS1 had already previously
been extensively investigated. The central energy source is estimated
to be an O6 star with a mass of $\sim$30\,M$_{\odot}$ and a luminosity
of $\sim 8\times 10^4$\,L$_{\odot}$ (e.g.,
\citealt{willner1976,campbell1984,gaume1995b,pestalozzi2004,sandell2004,reid2005}).
While \citet{campbell1984b}, \citet{gaume1995b} and
\citet{sandell2009} report an ionized jet in the north-south direction,
\citet{minier2000} and \citet{pestalozzi2004,pestalozzi2009} present
the detection of CH$_3$OH class II masers at 6.7 and 12.2\,GHz, which is
indicative of an accretion disk perpendicular to the outflow. Partly
different interpretations arise from mid-infrared continuum imaging
\citep{debuizer2005c} that detect a disk-like structure in the
northeast-southwestern direction perpendicular to the molecular
outflow \citep{keto1991,davis1998}. Figure \ref{cont} shows the
various outflow/disk axes discussed in the literature so far.  Setting
high-spatial-resolution near-infrared speckle images of the region
into context with the existing data, \citet{kraus2006} propose
precession of the underlying disk-jet structure as the main reason for
observing different axis orientations. Furthermore,
\citet{hoffman2003} report high-spatial-resolution observations of
the rare H$_2$CO maser emission that is also consistent with a very
young disk candidate (cf.~\citealt{araya2007b}). In addition to the
proposed CH$_3$OH maser disk, \citet{minier2000} identified a few
additional CH$_3$OH maser positions, several of them approximately in
the east-west direction but one also in the south of the core
(Fig.~\ref{cont}). All these additional maser components are
blue-shifted with respect to the $v_{\rm{lsr}}$ and the southern
position is said to be associated with the ionized north-south
jet \citep{minier2001}.  \citet{qiu2011} observed red-shifted
absorption toward NGC7538IRS, which they interpret as infall motions
of the dense gas.

Resolving this region with the PdBI in the 1.3\,mm continuum emission,
as well as several spectral lines at $\sim$800\,AU resolution, the
source still remains one compact continuum core without evidence of
significant fragmentation \citep{beuther2012c}. Furthermore, in the
dense gas spectral lines, we identified a velocity gradient in the
northeast-southwest direction that is consistent with the mid-infrared
disk hypothesis of \citet{debuizer2005c}. However, the velocity
profiles are strongly affected by absorption against the strong
continuum, which made interpretating of velocity structures from a
potential disk almost impossible. Nevertheless, the absorption
profiles allowed us to estimate approximate mass infall rates on the
order of $10^{-2}$\,M$_{\odot}$\,yr$^{-1}$.

Based on these exciting results, we now aim to resolve and study this
promising massive disk candidate at the highest spatial resolution
possible in pre- and early-ALMA (Atacama Large Millimeter Array) time,
as well as in higher excited lines that may not be as affected by the
absorption than the lines of the previous studies. It should also be
noted that this important northern-sky star-forming region will never
be accessible with ALMA. Therefore, we observed NGC7538\,IRS1 at the
shortest wavelengths accessible with the PdBI at 843\,$\mu$m in its
most extended configuration. These observations resolve the submm
continuum and spectral line emission (HCN$(4-3)v_2=1$,
CH$_3$OH$(15_{1,14}-15_{0,15})$ and HCO$^+(4-3)$) at previously
unaccessible $0.2''\times 0.17''$ spatial resolution corresponding to
linear resolution elements of $\sim$500\,AU.  Questions we are
addressing with these observations are: What are the fragmentation
properties of the innermost core region? What are the rotational
properties of that entity?  How does the innermost outflow structure
relate to the rotational structure?

\section{Observations} 
\label{obs}

NGC7538\,IRS1 was observed, together with NGC7538S, in a shared-track mode
in A configuration on February 27, 2012. The phase center of
NGC7538\,IRS1 was R.A (J2000.0)  23h\,13m\,45.360s, Dec.~(J2000.0)
61$^o$\,28$'$\,10.55$''$. Phase calibration was conducted with
regularly interleaved observations of the quasars 0059+581, 0016+731,
and 2320+506. The bandpass and flux were calibrated with observations
of 3C279. The absolute flux level is estimated to 20\% accuracy.  The
continuum emission was extracted from apparently line-free broad band
data obtained with the WIDEX correlator with four units and two
polarizations covering the frequency range from 354.16 to 357.77\,GHz.
The $1\sigma$ continuum rms for NGC7538\,IRS1 is 50\,mJy\,beam$^{-1}$.
To extract kinematic information, we put several high-spectral
resolution units with a nominal resolution of 0.312\,MHz or
0.26\,km\,s$^{-1}$ into the bandpass covering the spectral lines with
upper level energies $E_u/k$ between 43 and 1050\,K (Table
\ref{linelist}).  The spectral line rms for 1.0\,km\,s$^{-1}$ wide
spectral channels measured in emission-free channels, as well as
channels with emission, varies between 30 and 55\,mJy\,beam$^{-1}$.
The $v_{\rm{lsr}}$ is $\sim -57.3$\,km\,s$^{-1}$ (Gerner et al.~in
prep., \citealt{vandertak2000,sandell2010}).  The data were inverted
with a ``robust'' weighting scheme and cleaned with the Clark
algorithm. The synthesized beam of the final continuum and line data
is $\sim 0.2''\times 0.17''$ (PA 93$^{\circ}$). While NGC7538\,IRS1 is
extraordinarily strong in line and continuum emission, NGC7538S is
much weaker -- it was not detected in the spectral lines -- and even
imaging of the continuum was difficult. Therefore, we do not discuss
the NGC7538S data.

\begin{table}[htb]
\caption{Observed spectral lines}
\begin{tabular}{lrr}
\hline \hline
Freq. & Mol. & $E_u/k$ \\
(GHz) &       &  (K) \\
\hline
356.007152  & CH$_3$OH$(15_{1,14}-15_{0,15})$ & 278 \\
356.2555682 & HCN$(4-3)v_2=1$ & 1050 \\
356.734242  & HCO$^+(4-3)$ & 43 \\
\hline \hline
\end{tabular}
~\\
\label{linelist}
\end{table}

\section{Results and discussion}

\subsection{Fragmentation of the innermost core in the 843\,$\mu$m continuum emission}
\label{continuum}

Figure \ref{cont} presents the highest angular-resolution submm
continuum image (843\,$\mu$m or 356\,GHz) that has, to the authors'
knowledge, ever been obtained from a high-mass protostar and disk
region.  The synthesized beam of $0.2''\times 0.17''$ corresponds at
the given distance of 2.7\,kpc to a spatial resolution element of
approximately 500\,AU. For the first, therefore, time we resolve the
structures where not only the massive dense cores but also the
proposed embedded high-mass disks are expected to fragment (e.g.,
\citealt{krumholz2007a}).

\begin{figure*}[ht!] 
\includegraphics[width=0.98\textwidth]{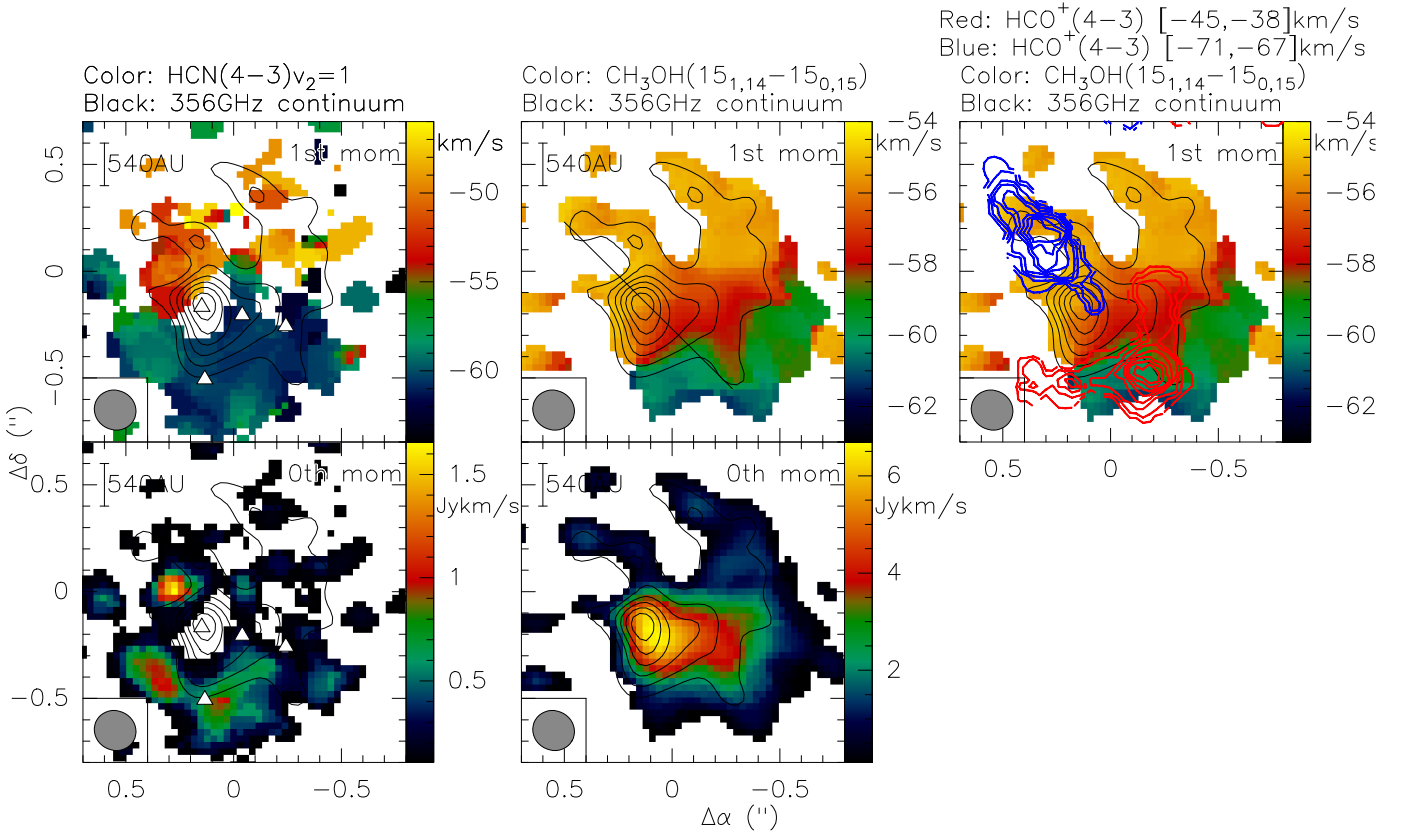}
\caption{The color scales in the left panels present the 0th moment
  (bottom, integrated intensity) and 1st moment (top, intensity
  weighted velocity) maps of the HCN(4--3)$v_2=1$ emission. In the
  middle and right panels, the color-scale shows the 0th and 1st
  moment maps (bottom and top, respectively) of
  CH$_3$OH$(15_{1,14}-15_{0,15})$. Black contours in all panels
  outline the dust continuum emission (Fig.~\ref{cont}) starting at
  the $3\sigma$ level and continuing in $2\sigma$ steps. The blue and
  red contours in the right panel show the blue- and red-shifted
  HCO$^+(4-3)$ emission integrated over the velocity regimes written
  above the panel. Contouring is done from 10 to 90\% of the
  respective peak emission. The white triangles present the additional
  maser positions by \citet{minier2000}, and a beam and scale bar are
  presented as well. The CH$_3$OH 1st moment map also marks the axis
  along which the position-velocity cuts in Fig.~\ref{pv} are
  conducted.}
\label{moments}
\end{figure*}

While the lower resolution 1.36\,mm image in \citet{beuther2012c}
consisted mainly of an unresolved point source with only very little
extended emission at low intensities, the new 843\,$\mu$m data now
reveal substructure within the inner $\sim$3000\,AU of the region. In
addition to spatially resolving the central peak into an elongated
structure with a subpeak about $0.3''$ ($\sim 900$\,AU) north of the
main peak (Fig.~\ref{cont}), there appears to be a secondary structure
toward the northwest of the main peak. While the arm-like structure is
at the 3-4$\sigma$ level, the northwestern peak of that arm (offset
$-0.11''/0.36''$ or $\sim 1300$\,AU, labeled submmNW in
Fig.~\ref{cont}) is above $5\sigma$. The lowest $4\sigma$ contour of
the previous 1.36\,mm data ($1\sigma\sim 29$\,mJy\,beam$^{-1}$,
\citealt{beuther2012c}) shows only a weak extension in that direction.
The peak 843\,$\mu$m flux from submmNW is
$\sim$0.27\,mJy\,beam$^{-1}$, and assuming a $\nu^4$ flux dependence
at (sub)mm wavelength, which translates into a 1.36\,mm flux of
$\sim$40\,mJy\,beam$^{-1}$. This is below the $2\sigma$ level of the
previous 1.36\,mm data, so it explains the non-detection at the longer
wavelengths. The detection of fragmenting substructure in the dense
core/disk at the shorter wavelengths shows the power of this
wavelength range.

The peak and integrated (above $3\sigma$ level) 843\,$\mu$m fluxes
extracted from Fig.~\ref{cont} are 0.74\,Jy\,beam$^{-1}$ and 3.38\,Jy,
respectively. At the given spatial resolution, the peak and 3$\sigma$
fluxes of 0.74 and 0.15\,Jy\,beam$^{-1}$ correspond to brightness
temperatures from the peak to the edge between 219 and 51\,K.
\citet{reid2005} report an 850\,$\mu$m single-dish continuum peak flux
of 18.1\,Jy\,beam$^{-1}$, which implies that our interferometer
observations filter out approximately 80\% of the total flux. To
calculate masses and column densities, it is important to estimate the
free-free contribution to the total flux measured at the given
wavelengths. While the free-free emission has approximately a $\nu^2$
dependence in the optically thick regime, it turns almost flat
($\nu^{-0.1}$) in the optically thin regime. As a result, if one knows the
turnover frequency between the two regimes such an estimate is
relatively easy. However, that turnover frequency has not been determined
yet for NGC7538\,IRS1. \citet{sandell2009} compiled the cm wavelength
fluxes and find rising fluxes up to 43.4\,GHz where they measure
a flux of $\sim$430\,mJy toward the central elongated but still
compact core (size $0.2''\times 0.14''$, smaller than our synthesized
beam). This value can be considered as a lower limit of the free-free
contribution even at submm wavelengths.  But because the spectral
energy distribution has not reached the optically thin regime at that
wavelength yet, giving an upper limit is difficult. In our previous
1.36\,mm study we assumed 1\,Jy as a free-free flux contribution;
however, that is obviously too high because we only measure
0.74\,Jy\,beam$^{-1}$ peak flux in our new submm PdBI data.  As an
approximation, we assume 0.5\,Jy as a free-free flux contribution at
843\,$\mu$m, which results in a dust-related peak and integrated fluxes
at that wavelength of 0.24\,Jy\,beam$^{-1}$ and 2.88\,Jy,
respectively. Interestingly, after correcting for the free-free
emission, the dust related 843\,$\mu$m peak flux from the central peak
is very close to that measured from the northern peak (submmN) with
$\sim$0.26\,mJy\,beam$^{-1}$ and the northwestern extension peak
(submmNW) with $\sim$0.27\,mJy\,beam$^{-1}$. While even submmN and
submmNW may make free-free contributions to the measured flux, based
on the published cm data, this potential contribution should be very
small \citep{sandell2009}.

Assuming optically thin emission of dust at submm wavelength at an
approximate temperature of 245\,K \citep{qiu2011} with a dust opacity
$\kappa\sim 1.8$\,cm$^2$\,g$^{-1}$ (extrapolated from
\citealt{ossenkopf1994} for densities of $10^6$\,cm$^{-3}$ and dust with
thin ice mantles) and a gas-to-dust mass ratio of 186
\citep{jenkins2004,draine2007}, the integrated flux corresponds to a
total mass of the submm structure of $\sim$11\,M$_{\odot}$. Using the
peak fluxes to estimate the mass contributions of resolved fragments
(central submmN and submmNW), they have similar values with 0.9, 1.1,
and 1.0\,M$_{\odot}$, respectively. 
The gas column densities toward the central peak, as well as submmN and
submmNW, are approximately the same with $0.9\times 10^{25}$,
$1.0\times 10^{25}$, and $1.0\times 10^{25}$\,cm$^{-2}$, respectively.
This corresponds to very high visual extinction values around
$10^5$\,mag. For comparison, assuming ISM dust properties
\citep{mathis1977} in the classical \citet{hildebrand1983} picture the
masses and column densities would be higher with values of
$\sim$25\,M$_{\odot}$, $2.0\times 10^{25}$, $2.1\times 10^{25}$, and
$2.2\times 10^{25}$\,cm$^{-2}$, respectively.  Independent of the
assumed dust properties, the detection of a central infrared source
(e.g., \citealt{puga2010}), in spite of such extremely high gas
column density and extinction values, indicates that the outflow is
likely to be aligned not to far off the line-of-sight excavating a
cavity through which we can peer into the center (see section
\ref{outflow}).

In addition to the fragmenting structures on 1000\,AU scales, the
extremely high spatial resolution of this continuum map also reveals
an elongation of the central peak around NGC7538\,IRS1. At the
$>10\sigma$ level, the 843\,$\mu$m emission appears elongated along an
axis approximately inbetween the north-south jet axis and the
northeast-southwest mid-infrared and molecular disk axis marked in
Figure \ref{cont}. This orientation of the elongation reflects the
likely situation that the submm emission receives contributions from the
assumed underlying dusty disk, as well as from the ionized jet.
However, we cannot exclude that the elongation could also be due to an
unresolved binary. \citet{kraus2006} do not find any equal brightness
binary at their spatial resolution limit of 70\,mas, but their
proposed precession model requires a very close and unresolved binary
component. Furthermore, the data reveal elongated and more extended
emission $\sim 0.3''$ east of the main peak. While this emission can
be part of the smoother envelope emission, it may also host another
unresolved subsource in the region.

The fragmenting core structure is consistent with - if not a
prerequisite for - the observed high degree of multiplicity and
Trapezia systems in high-mass star formation (e.g.,
\citealt{zinnecker2007}). While fragmentation can also be suppressed,
for example, by magnetic fields (e.g., \citealt{commercon2011}),
recent hydrodynamical simulations of collapsing gas clumps predict the
fragmentation of the rotating cores/disks and even competitive
accretion between the different subfragments within the disk
\citep{krumholz2009,peters2010b}.  The model-predicted fragmentation
scales of the inner core/disk structures range between several 100 and
a few 1000\,AU, resembling the observed structures fairly closely.

\begin{figure}[ht!] 
\includegraphics[width=0.49\textwidth]{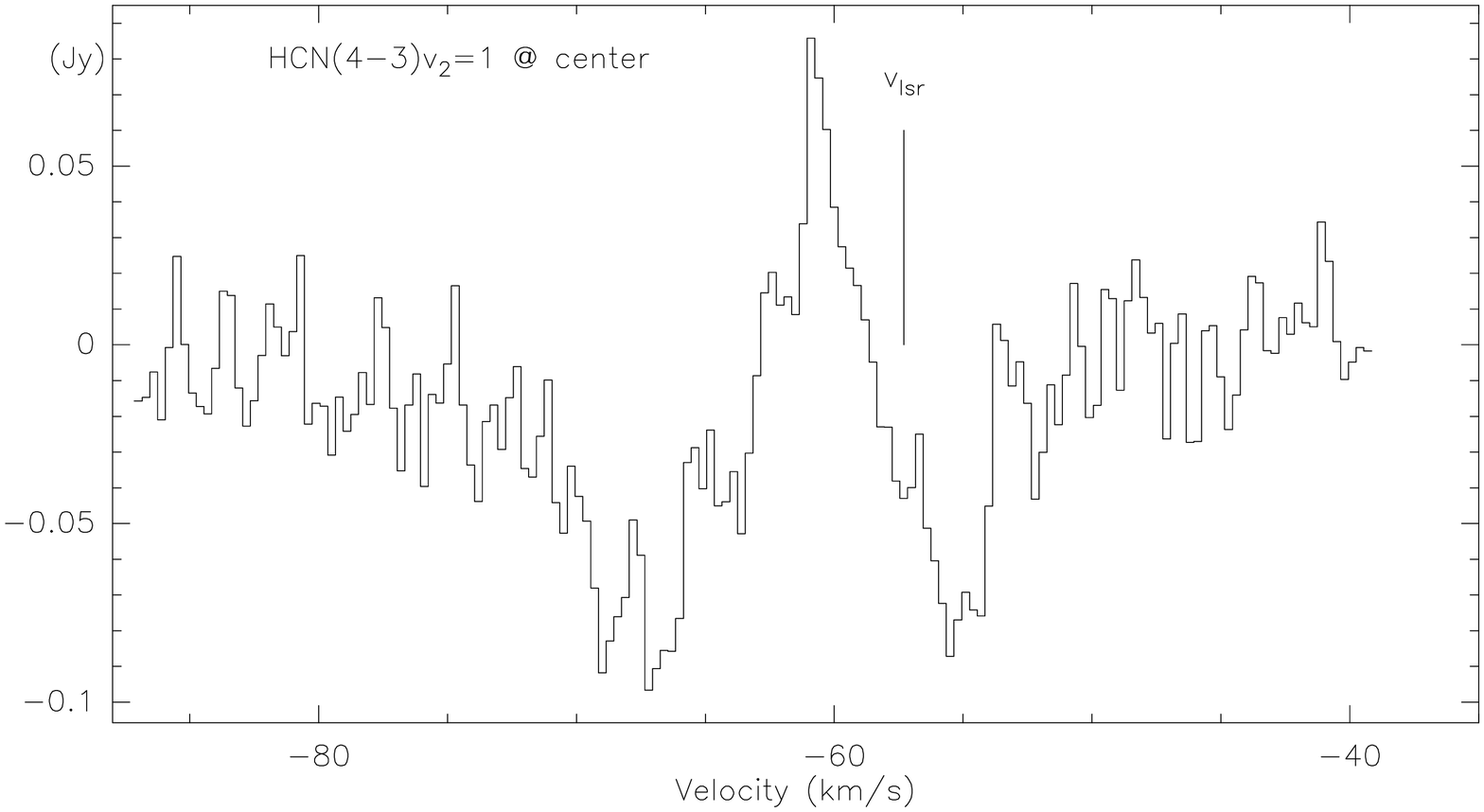}
\includegraphics[width=0.49\textwidth]{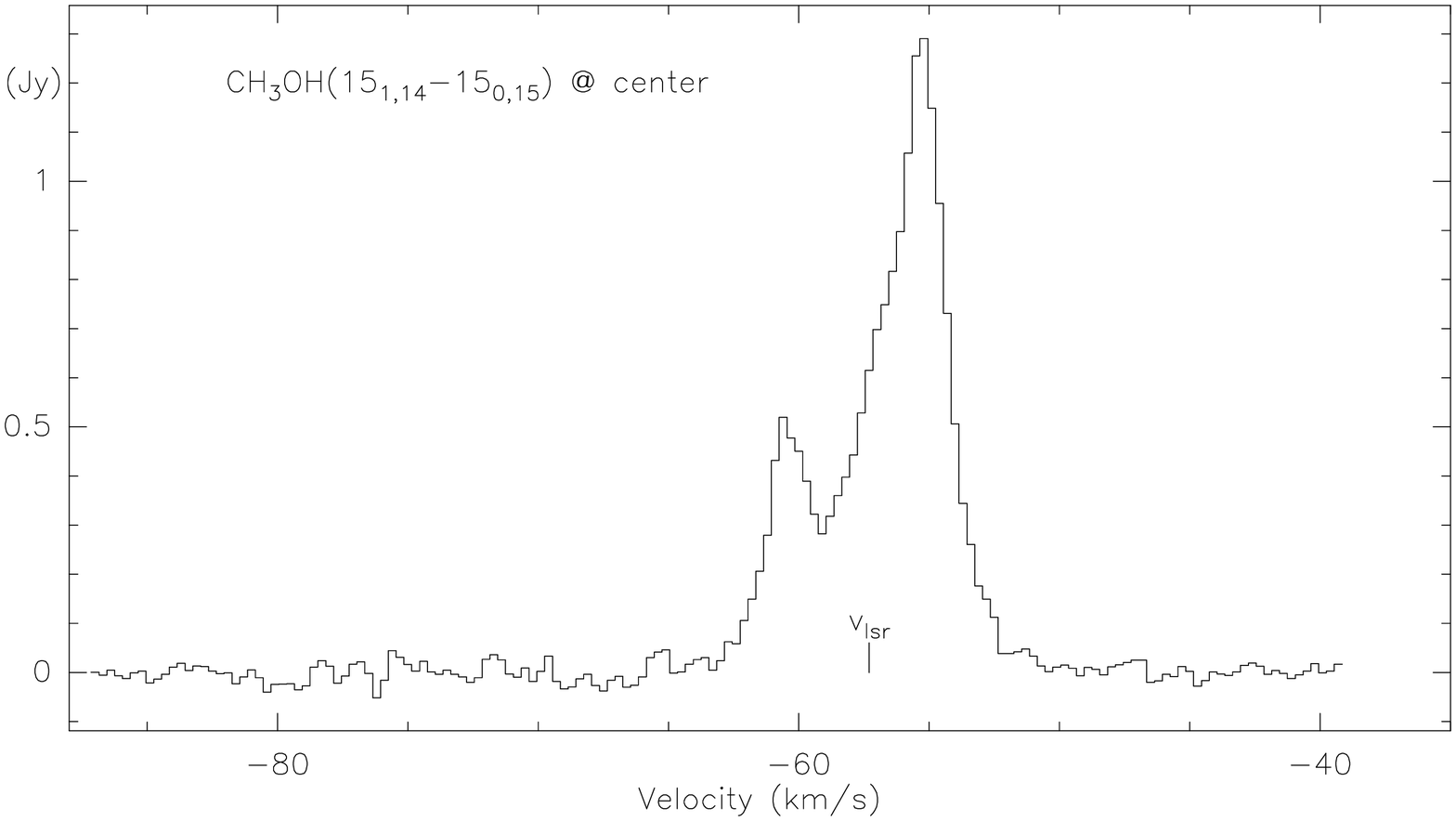}
\includegraphics[width=0.49\textwidth]{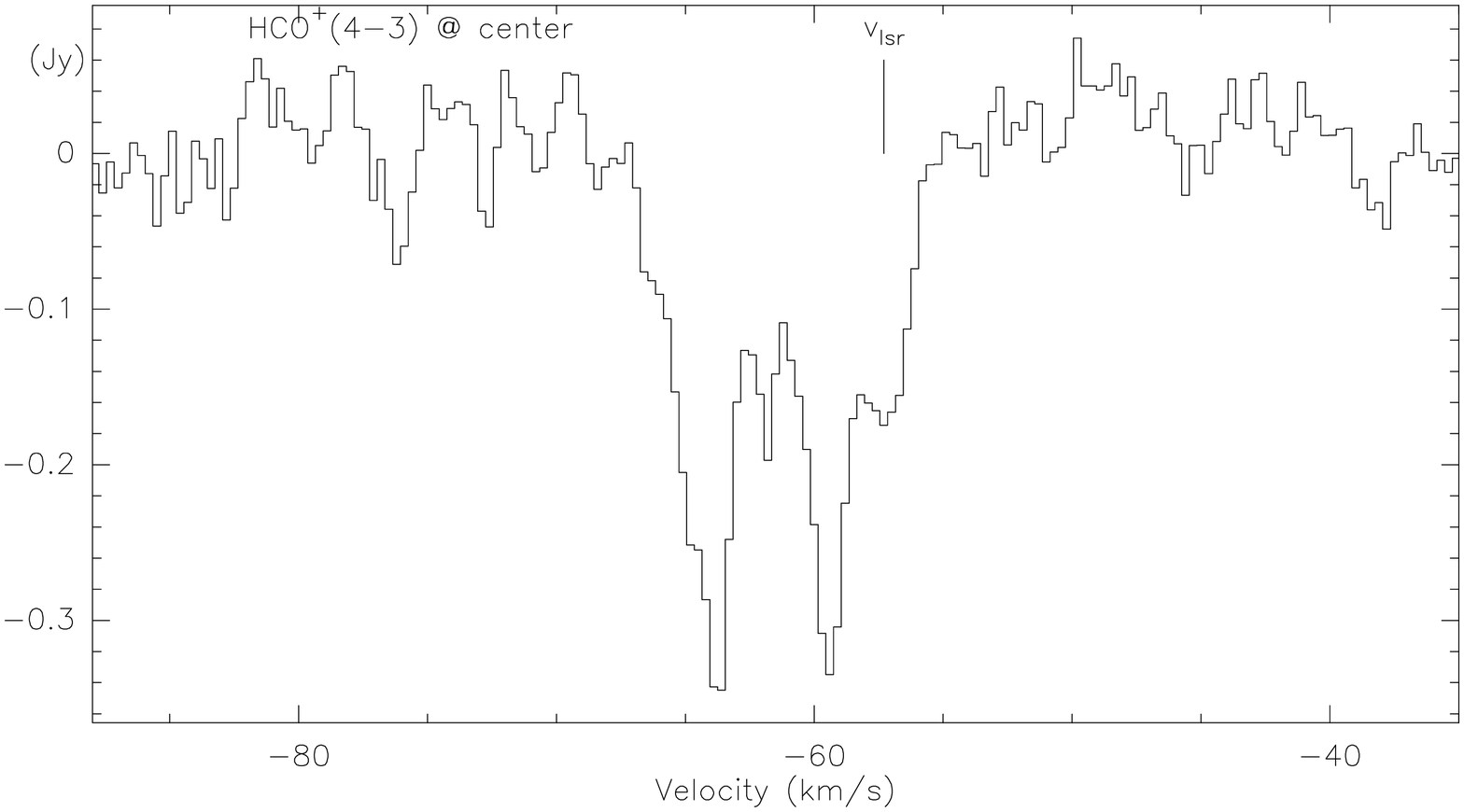}
\includegraphics[width=0.49\textwidth]{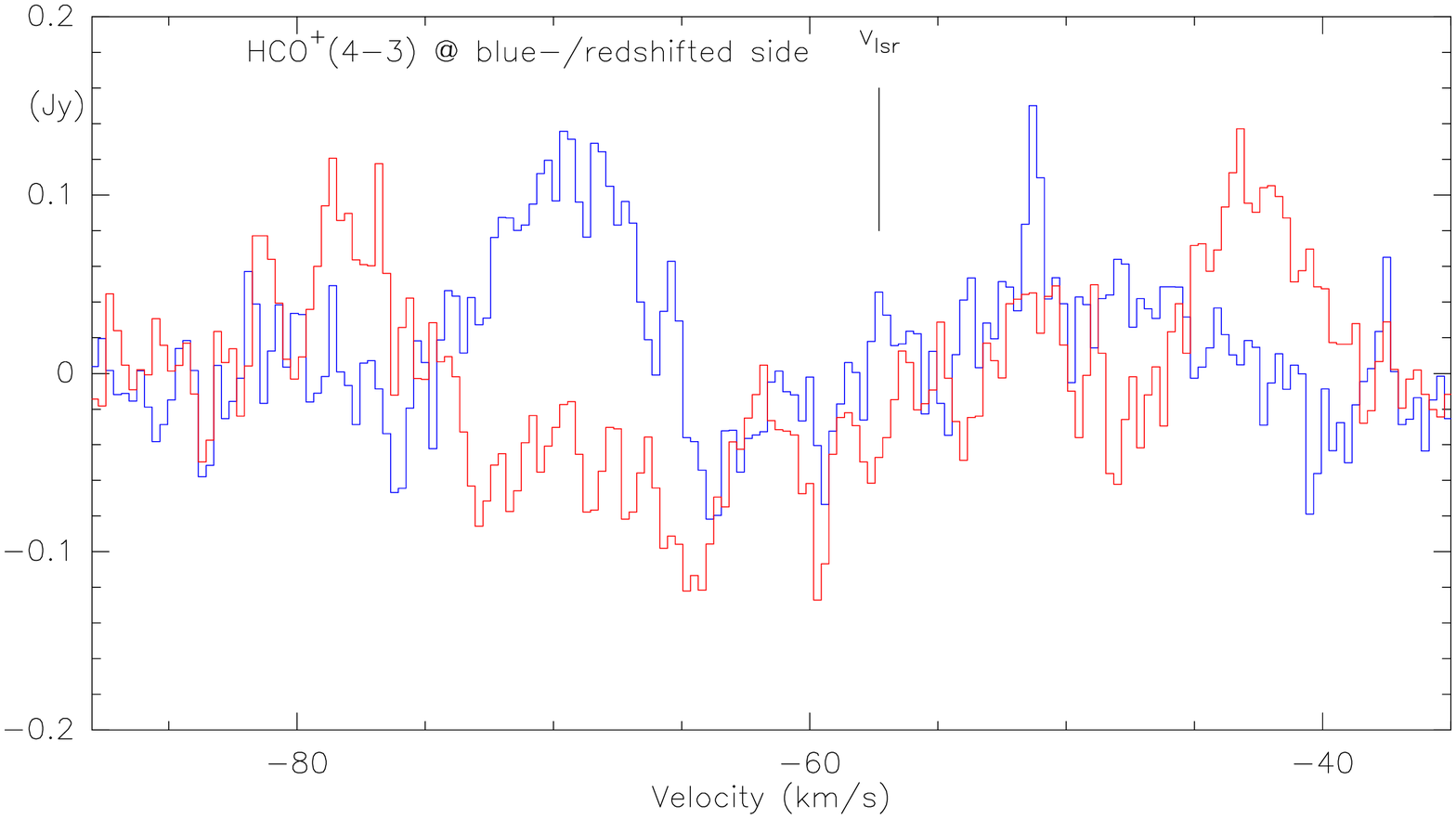}
\caption{The top three continuum-subtracted spectra are extracted for
  HCN, CH$_3$OH and HCO$^+$ toward the central peak position (offset
  $0.1''/-0.2''$). The bottom spectrum shows the HCO$^+$ spectra from
  the blue- and red-shifted outflow peak positions presented in
  Fig.~\ref{moments} (offsets positions $0.33''/0.18''$ and
  $-0.18''/-0.48''$, respectively).}
\label{spectra}
\end{figure}

\subsection{Gas kinematics}

The spectral line data also allow us to investigate the gas kinematics
of the different components in more detail. Figure \ref{moments} gives
an overview of the HCN$(4-3)v_2=1)$, CH$_3$OH$(15_{1,14}-15_{0,15})$,
and HCO$^+(4-3)$ results. While we show for HCN$(4-3)v_2=1$ and
CH$_3$OH$(15_{1,14}-15_{0,15})$ zeroth (integrated intensity) and
first moment maps (intensity-weighted velocities), the blue- and
red-shifted high-velocity gas is presented for HCO$^+(4-3)$. Figure
\ref{spectra} additionally shows the spectra of all three lines
extracted toward the continuum peak, as well as the HCO$^+(4-3)$
spectra toward the blue- and red-shifted outflow peak positions.

While both high-density tracers -- HCN$(4-3)v_2=1$ and
CH$_3$OH$(15_{1,14}-15_{0,15})$ -- clearly show the
northeast-southwest velocity gradient of the underlying rotating
core/disk structure on scales of approximately 3000\,AU, it is
interesting that the highest excited HCN$(4-3)v_2=1$ line with an
upper energy level $E_u/k$ of 1050\,K still exhibits the central
absorption features. Considering the peak brightness temperature of
the continuum emission of 219\,K, this implies that the beam-averaged
excitation temperature of the spectral line has to be lower than that.
At the high densities within the core, the same applies to the
beam-averaged gas kinetic temperatures. The $E_u/k=278$\,K
CH$_3$OH$(15_{1,14}-15_{0,15})$ line is among the few lines that do
not show any absorption at high spatial resolution.  From the suite of
lines presented in \citet{beuther2012c}, that covers a broad range of
energy levels between 21 and 326\,K, only the optically thin
methyl-cyanide isotopologue CH$_3^{13}$CN did not show any prominent
absorption feature. Since the CH$_3$OH$(15_{1,14}-15_{0,15})$ line is
very strong and probably not optically thin, it does not seem to be an
optical depth effect here.  Why this line does not show up in
absorption is not exactly clear yet. However, it is known that many
CH$_3$OH transitions are potential masers. One condition for the
inversion population for masers is a negative excitation temperature.
On the way to inversion population, the excitation temperature first
asymptotically rises to high $T_{\rm{ex}}$ (e.g.,
\citealt{stahler2005}, Fig.~14.6) before it inverts and becomes
negative. Hence, even in the non-masing state, a molecule like
CH$_3$OH may more easily exhibit high excitation temperatures and thus
less absorption than most other molecules.  Independent of that, the
CH$_3$OH emission line is very useful for studying the rotational
properties of the core without absorption artifacts.

\subsubsection{Rotational properties}

\begin{figure*}[ht] 
\includegraphics[width=0.98\textwidth]{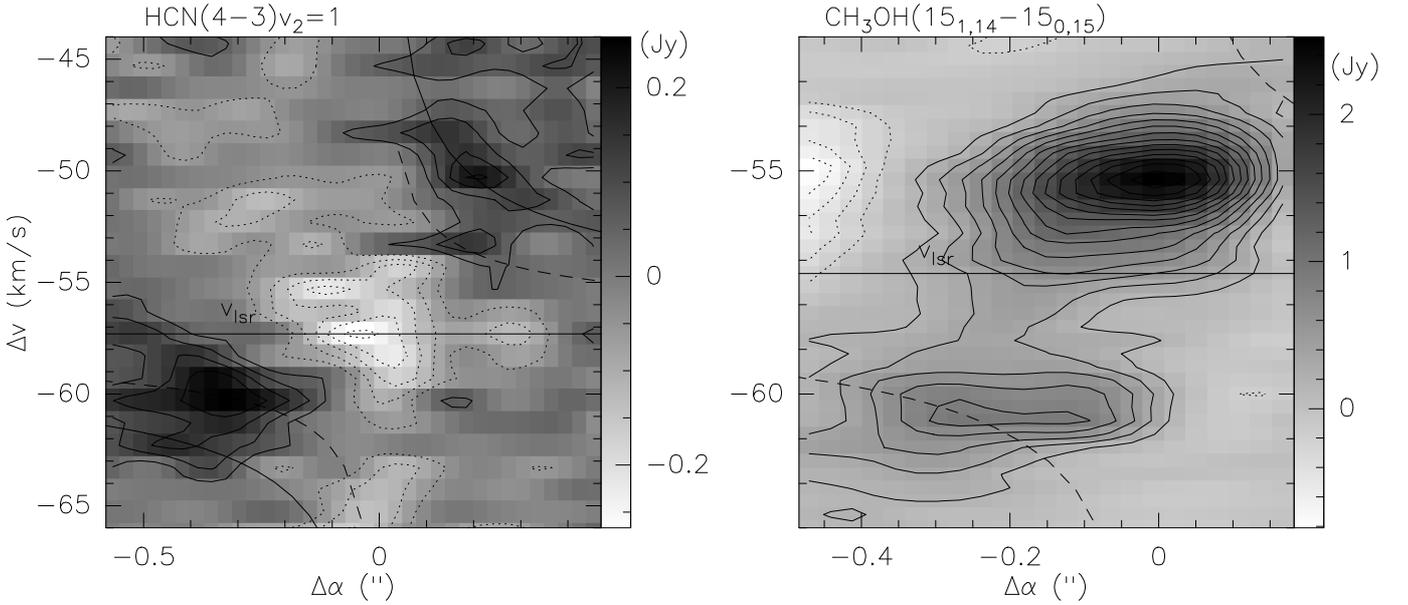}
\caption{Position-velocity diagrams for HCN(4--3)$v_2=1$ (left)
  CH$_3$OH$(15_{1,14}-15_{0,15})$ (right) along the
  northeast-southwest, mid-infrared and molecular axis shown in
  Fig.~\ref{cont}.  The full and dashed lines mark the rotation curves
  for embedded 30 and 8\,M$_{\odot}$ objects, respectively. The
  $v_{\rm{lsr}}$ is shown as well.}
\label{pv}
\end{figure*}

While a rotating inner envelope and/or disk structure is already
evident from the first moment map (Fig.~\ref{moments}), to investigate
whether a Keplerian accretion disk can be identified with the new
data, Figure \ref{pv} presents position-velocity cuts for
HCN$(4-3)v_2=1)$ and CH$_3$OH$(15_{1,14}-15_{0,15})$ along the
northeast-southwest axis most likely representing the orientation of
the proposed embedded disk. This axis is not exactly aligned with the
extension of the continuum emission (Sec.~\ref{continuum}), but it is
the axis of the strongest velocity gradient (Figs.~\ref{cont} \&
\ref{moments}).  Although we identify a clear velocity gradient across
the core, thereby confirming its general rotation, the
position-velocity diagram is not consistent with Keplerian rotation,
but the data instead exhibit sub-Keplerian motions of the gas. To
visualize this, Fig.~\ref{pv} shows the position-velocity signatures
one would expect for a central 30\,M$_{\odot}$ and 8\,M$_{\odot}$
star. Obviously, none of these rotation curves fits the observations,
which implies that we still miss clear signatures of a Keplerian
accretion disk around genuine high-mass young stellar objects.

While Keplerian disks have been identified and studied intensively for
low-mass TTauri stars (e.g., \citealt{simon2000}), Keplerian rotation
has been identified only in a few cases around early B-type stars
(e.g., AFGL490 or IRAS\,20126, \citealt{schreyer2002,cesaroni2005}).
However, in the high-mass regime mainly larger scale rotating toroids
have been found (e.g., \citealt{cesaroni2007,fallscheer2009}).
Recently, a large PdBI survey of several low-mass class 0 sources did
not identify any Keplerian disk on scales larger than 50\,AU
(\citealt{maury2010}, Maury et al.~in prep.). This implies that the
real Keplerian disks around very young low-mass protostars have to be
very small in size.  Extrapolating this picture to the high-mass
sources, we have also not found Keplerian disks on scales $> 500$\,AU.
One should keep in mind that the rotating envelope mass of around
11\,M$_{\odot}$ is a significant fraction of the reported stellar mass
of 30\,M$_{\odot}$ (which is an upper limit because the mass estimates
based on the luminosity do not properly take the potential accretion
luminosity into account). Since a requirement for Keplerian rotation
is that the disk mass is negligible compared to the mass of the
central object, real Keplerian signatures may not even be expected on
these larger scales.  Since in both the low-mass and high-mass cases,
collimated outflows are known that indicate stable disks for their
launching, in the high-mass case we also have to achieve better
spatial resolution and search for these entities on still smaller
scales. Small inner Keplerian disks are also consistent with
near-infrared spectral line observation results (e.g.,
\citealt{thi2005,wheelwright2010,ilee2013}.) In principle, one could
fathom such a small inner accretion disk also in the ionized gas,
however, high-spatial-resolution recombination line studies of this
region also exhibit the north-south jet-structure with very broad
recombination lines \citep{gaume1995b}. These are more
straightforwardly interpreted in the framework of an outflow/wind
model for the ionized gas.

It is interesting to note that the additional blue-shifted CH$_3$OH
maser components that are not part of the proposed Keplerian disk but
found a bit offset (Figs.~\ref{cont} and \ref{moments},
\citealt{minier2000,minier2001}) are spatially and velocity-wise
correlated with the rotating gas observed in HCN and CH$_3$OH and less
with the inverse velocity-structure seen in HCO$^+$. This strongly
indicates that the masers are not part of the outflow but of the inner
rotating core.

\subsubsection{Infall rates}
\label{infall}

The HCN$(4-3)v_2=1$ spectrum extracted toward the submm peak positions
exhibits clear blue- and red-shifted absorption signatures.
\citet{knez2009} found similar spectral absorption signatures in
spectral line observations at mid-infrared wavelengths, and they
interpret their data in the framework of absorption from a nearly
edge-on disk. We think that the expected underlying embedded
disk-structure is currently unlikely to be too close to edge-on but
has to have a considerable inclination angle mainly for two reasons:
(1) If a flattened disk-like structure were fully edge-on, the central
source would be highly extincted and no or barely any emission would
be expected to escape at near-infrared wavelengths. In contrast to
that picture, near-infrared observations detect a central source
(e.g., \citealt{puga2010}). While the infrared emission could also in
principle be due to scattered light from the outflow cone (e.g.,
\citealt{kraus2006}), a simpler way to explain such a detection in the
presence of high column densities and extinctions as reported in
section \ref{continuum} is that one observes the region at least at
medium inclination through the cleared-out outflow cavity.  (2)
Furthermore, the velocity structure observed in HCO$^+$(3--2) (see
section \ref{outflow}) indicates that we have to look at least to some
degree into the cavity of the outflow. An additional aspect of the
whole region to be considered is the strong precession of the
disk-outflow system proposed by \citet{kraus2006}. While such
precession cannot only change the projected angle of the outflow on
the plane of the sky, it would also change the inclination to the
observer.  Therefore, in this precession model it is possible that the
outflow cone is closer to the line-of-sight during different
time steps in the evolution of the system.

Therefore, we favor a geometry with the outflow orientation being
closer to the line of sight where any flattened disk-like structure
has to be closer to face-on. Deriving an exact inclination angle is
not possible with the given data. In this framework, the blue-shifted
absorption should stem from outflowing gas, whereas the red-shifted
absorption indicates infalling gas.  Observing this infall in a line
with $E_u/k$ higher than 1000\,K implies that the infall continues
down to the innermost region around the star.  Although it is unclear
what fraction of the gas will finally be accreted, infall rates
estimated from these signatures should give an upper limit for the
actual accretion rates.

Assuming a spherical infall geometry, one can estimate mass infall
rates via $\dot{M}_{\rm{in}} = 4\pi r^2 \rho v_{\rm{in}}$ where
$\dot{M}_{\rm{in}}$ and $v_{\rm{in}}$ are the infall rate and infall
velocity, and $r$ and $\rho$ the core radius and density (see also
\citealt{qiu2011,beuther2012c}). We estimate the radius as $r=250$\,AU,
which is half our spatial resolution limit, corresponding well to the
area where absorption is present (Figure \ref{moments}, left
panels). The estimated density $\rho\sim 1.2\times 10^9$\,cm$^{-3}$
results from assuming that the peak column density derived above is
distributed along the line-of-sight at our spatial-resolution limit,
i.e., 500\,AU. The infall velocity can be estimated to
3.2\,km\,s$^{-1}$, which corresponds to the difference between the most
red-shifted absorption at $\sim -54.1$\,km\,s$^{-1}$, and the
$v_{\rm{lsr}}\sim -57.3$\,km\,s$^{-1}$. With these numbers, we derive
an infall rate of $\dot{M}_{\rm{in}} \sim 3.6\times
10^{-3}$\,M$_{\odot}$\,yr$^{-1}$.
This estimate is approximately a factor 20 lower than the infall rates
derived previously from the 1.36\,mm line data \citep{beuther2012b}.
The main difference stems from the higher spatial resolution of the
new data that enters the equation for $\dot{M}_{\rm{in}}$ through
$r^2$. Adding in the higher upper level energy of the HCN$(4-3)v_2=1$
line, the new value should resemble the actual accretion rates better
than the previous estimate.  To additionally improve our 1D spherical
infall estimate into a 2D disk geometry, we follow the arguments given
in \citet{beuther2012b}: Considering that the accretion does not occur
in a spherical mode over $4\pi$ but rather along a flattened disk
structure with a solid angle of $\Omega$, the actual disk infall rates
$\dot{M}_{\rm{disk,in}}$ should scale like
$\dot{M}_{\rm{disk,in}}=\frac{\Omega}{4\pi}\times \dot{M}_{\rm{in}}$.
Based on the simulations by \citet{kuiper2012} and R.~Kuiper
(priv.~comm.), the outflow covers approximately an 120\,degree opening
angle, and the disk 60\,degree (to be doubled for the north-south
symmetry). Since the opening angle does not scale linearly with the
surface element, full integration results in $\sim$50\% or $\sim 2\pi$
of the sphere being covered by the disk.  This then results in disk
infall rates of $\dot{M}_{\rm{disk,in}}\sim 1.8\times 10^{-3}$
\,M$_{\odot}$\,yr$^{-1}$, which is still very high and in the regime
of accretion rates required to form high-mass stars (e.g.,
\citealt{wolfire1987,mckee2003}).

\subsubsection{Outflow properties}
\label{outflow}

The blue- and red-shifted outflow signatures in the H CO$^+$(-3--2)
line in the northeast-southwest direction come as a surprise at first
sight. They appear aligned with the northeast-southwest axis of the
rotating structure, almost perpendicular to the CO and mid-infrared
outflow and still at a considerable angle to the ionized jet. The
facts that the HCO$^+$ velocities are significantly higher than the
rotational velocities measured in this direction by the CH$_3$OH and
HCN lines, as well as the inversion of blue- and red-shifted gas with
respect to the rotational structure, strongly indicate that the
HCO$^+$ emission traces different gas than do the other presented
lines.

Additional information can be extracted from the spectra taken toward
the blue- and red-shifted peak positions (Fig.~\ref{spectra}). These
spectra clearly show the blue- and red-shifted line wings, but both
spectra exhibit additional emission components. The spectrum toward
the blue-shifted peak position has another component at $\sim
-51.3$\,km\,s$^{-1}$ that is red-shifted with respect to the
$v_{\rm{lsr}}$. Correspondingly, the spectrum extracted from the
red-shifted peak exhibits an additional emission component at $\sim
-78.6$\,km\,s$^{-1}$ that is extremely blue-shifted compared to the
$v_{\rm{lsr}}$. Combining these two spectra, we see that both lobes
exhibit blue- and red-shifted emission. In an outflow scenario, this
indicates that the outflow is aligned almost along the line-of-sight
and that emission from both outflow lobes is found within the same
observational beams. The projected molecular HCO$^+$ outflow
geometry indicated by Fig.~\ref{moments} thus does not really reveal the
actual outflow geometry, but the outflow has to be aligned closer to
the line-of-sight. This agrees well with the fact that we see
infrared emission from a central object (e.g.,
\citealt{linz2009,puga2010}) from a region with such high column
densities, hence visual extinction (see section \ref{continuum})
that should prohibit any emission at short wavelengths. With an
outflow cone along the line-of-sight, the near-infrared emission can
easily escape and reach the observer (e.g.,
\citealt{hofner2001,linz2005}).

As already discussed in section \ref{infall}, the proposed precession
of the system has to be taken into account as well \citep{kraus2006}.
This implies that the inclination of the outflow cone with respect to
the line-of-sight can vary during different time steps in the
evolution of the system.

\section{Conclusions}
\label{conclusion}

These so far highest spatial-resolution submm observations of the
dense dust and thermal gas around a high-mass protostar reveals
several exciting results. For the first time, we have resolved
fragmentation properties of the rotating inner core with at least
three subsources within the inner 3000\,AU.  This is consistent with
the high degree of multiplicity found for high-mass stars. The region
also resembles structures found recently in hydrodynamical simulations
of collapsing gas clumps. In addition to the fragmentation results,
the data also clearly outline the importance of going to shorter
wavelengths to identify such fainter structures because the $\nu^4$
frequency dependence strongly outweights the poorer atmospheric
conditions at shorter wavelengths.

Identifying red-shifted absorption signatures of infall motions from a
high-excitation temperature ($E_u/k\sim 1050$\,K) tracer like the
HCN$(4-3)v_2=1$ line on scales $\leq 500$\,AU from the central
protostars allows us to estimate infall rates around $1.8\times
10^{-3}$\,M$_{\odot}$yr$^{-1}$. Although it is unclear what fraction
of the gas actually falls on the star, such inner core/disk infall
rates are among the best proxies of the actual accretion rates we can
get at that early evolutionary stage.

The outflow exhibits blue- and red-shifted emission on both sides of
the central object. This can only be explained by an outflow cone
opening almost along the line-of-sight, hence emitting blue- and
red-shifted gas within the same observational beam. Such a
line-of-sight outflow is also prerequisite for the simultaneous
detection of extremely high visual extinction values (on the order
$10^5$\,mag) and at the same time infrared emission from the central
object.

Collimated and jet-like outflows are usually considered to require
stable central, most likely Keplerian disks, and the search for these
objects is still open.  Although we clearly identify rotational
signatures of the core on $\sim$3000\,AU scales, even at our
resolution limit of 500\,AU we are not able to identify Keplerian
velocity signatures. The difficulty to identify disks during early
evolutionary stages has also recently been reported for low-mass class 0
sources (Maury et al.~in prep.).  Identifying and studying these low-
and high-mass accretion disks during the earliest evolutionary stages
remains one of the exciting topics in high-mass star formation for the
coming years.

While these observations are at the edge of the capabilities of
currently existing (sub)mm interferometers, they outline the immense
potential of ALMA and NOEMA in that field.

\begin{acknowledgements} 
  We would like to thank the IRAM staff, in particular Jan Martin Winters,
  for all support during the observation and data reduction process.
  Furthermore, we acknowledge the referee's report and comments
  that improved the paper.
\end{acknowledgements}


\begin{thebibliography}{64}
\expandafter\ifx\csname natexlab\endcsname\relax\def\natexlab#1{#1}\fi

\bibitem[{{Araya} {et~al.}(2007){Araya}, {Hofner}, {Goss}, {Linz}, {Kurtz}, \&
  {Olmi}}]{araya2007b}
{Araya}, E., {Hofner}, P., {Goss}, W.~M., {et~al.} 2007, \apjs, 170, 152

\bibitem[{{Argon} {et~al.}(2000){Argon}, {Reid}, \& {Menten}}]{argon2000}
{Argon}, A.~L., {Reid}, M.~J., \& {Menten}, K.~M. 2000, \apjs, 129, 159

\bibitem[{{Beltr{\'a}n} {et~al.}(2011){Beltr{\'a}n}, {Cesaroni}, {Neri}, \&
  {Codella}}]{beltran2011}
{Beltr{\'a}n}, M.~T., {Cesaroni}, R., {Neri}, R., \& {Codella}, C. 2011, \aap,
  525, A151

\bibitem[{{Beuther} {et~al.}(2007{\natexlab{a}}){Beuther}, {Churchwell},
  {McKee}, \& {Tan}}]{beuther2006b}
{Beuther}, H., {Churchwell}, E.~B., {McKee}, C.~F., \& {Tan}, J.~C.
  2007{\natexlab{a}}, in Protostars and Planets V, ed. B.~{Reipurth},
  D.~{Jewitt}, \& K.~{Keil}, 165--180

\bibitem[{{Beuther} {et~al.}(2012{\natexlab{a}}){Beuther}, {Linz}, \&
  {Henning}}]{beuther2012c}
{Beuther}, H., {Linz}, H., \& {Henning}, T. 2012{\natexlab{a}}, \aap, 543, A88

\bibitem[{{Beuther} {et~al.}(2002{\natexlab{a}}){Beuther}, {Schilke}, {Gueth},
  {McCaughrean}, {Andersen}, {Sridharan}, \& {Menten}}]{beuther2002d}
{Beuther}, H., {Schilke}, P., {Gueth}, F., {et~al.} 2002{\natexlab{a}}, \aap,
  387, 931

\bibitem[{{Beuther} {et~al.}(2002{\natexlab{b}}){Beuther}, {Schilke}, {Menten},
  {Motte}, {Sridharan}, \& {Wyrowski}}]{beuther2002a}
{Beuther}, H., {Schilke}, P., {Menten}, K.~M., {et~al.} 2002{\natexlab{b}},
  \apj, 566, 945

\bibitem[{{Beuther} {et~al.}(2012{\natexlab{b}}){Beuther}, {Tackenberg},
  {Linz}, {Henning}, {Schuller}, {Wyrowski}, {Schilke}, {Menten}, {Robitaille},
  {Walmsley}, {Bronfman}, {Motte}, {Nguyen-Luong}, \&
  {Bontemps}}]{beuther2012b}
{Beuther}, H., {Tackenberg}, J., {Linz}, H., {et~al.} 2012{\natexlab{b}}, \apj,
  747, 43

\bibitem[{{Beuther} {et~al.}(2007{\natexlab{b}}){Beuther}, {Zhang}, {Bergin},
  {Sridharan}, {Hunter}, \& {Leurini}}]{beuther2007d}
{Beuther}, H., {Zhang}, Q., {Bergin}, E.~A., {et~al.} 2007{\natexlab{b}}, \aap,
  468, 1045

\bibitem[{{Bontemps} {et~al.}(2010){Bontemps}, {Motte}, {Csengeri}, \&
  {Schneider}}]{bontemps2010}
{Bontemps}, S., {Motte}, F., {Csengeri}, T., \& {Schneider}, N. 2010, \aap,
  524, A18

\bibitem[{{Campbell}(1984)}]{campbell1984b}
{Campbell}, B. 1984, \apjl, 282, L27

\bibitem[{{Campbell} \& {Thompson}(1984)}]{campbell1984}
{Campbell}, B. \& {Thompson}, R.~I. 1984, \apj, 279, 650

\bibitem[{{Cesaroni} {et~al.}(2007){Cesaroni}, {Galli}, {Lodato}, {Walmsley},
  \& {Zhang}}]{cesaroni2007}
{Cesaroni}, R., {Galli}, D., {Lodato}, G., {Walmsley}, C.~M., \& {Zhang}, Q.
  2007, in Protostars and Planets V, ed. B.~{Reipurth}, D.~{Jewitt}, \&
  K.~{Keil}, 197--212

\bibitem[{{Cesaroni} {et~al.}(2005){Cesaroni}, {Neri}, {Olmi}, {Testi},
  {Walmsley}, \& {Hofner}}]{cesaroni2005}
{Cesaroni}, R., {Neri}, R., {Olmi}, L., {et~al.} 2005, \aap, 434, 1039

\bibitem[{{Commer{\c c}on} {et~al.}(2011){Commer{\c c}on}, {Hennebelle}, \&
  {Henning}}]{commercon2011}
{Commer{\c c}on}, B., {Hennebelle}, P., \& {Henning}, T. 2011, \apjl, 742, L9

\bibitem[{{Davis} {et~al.}(1998){Davis}, {Moriarty-Schieven}, {Eisl{\"o}ffel},
  {Hoare}, \& {Ray}}]{davis1998}
{Davis}, C.~J., {Moriarty-Schieven}, G., {Eisl{\"o}ffel}, J., {Hoare}, M.~G.,
  \& {Ray}, T.~P. 1998, \aj, 115, 1118

\bibitem[{{De Buizer} \& {Minier}(2005)}]{debuizer2005c}
{De Buizer}, J.~M. \& {Minier}, V. 2005, \apjl, 628, L151

\bibitem[{{De Buizer} {et~al.}(2005){De Buizer}, {Radomski}, {Telesco}, \&
  {Pi{\~n}a}}]{debuizer2005}
{De Buizer}, J.~M., {Radomski}, J.~T., {Telesco}, C.~M., \& {Pi{\~n}a}, R.~K.
  2005, \apjs, 156, 179

\bibitem[{{Draine} {et~al.}(2007){Draine}, {Dale}, {Bendo}, {Gordon}, {Smith},
  {Armus}, {Engelbracht}, {Helou}, {Kennicutt}, {Li}, {Roussel}, {Walter},
  {Calzetti}, {Moustakas}, {Murphy}, {Rieke}, {Bot}, {Hollenbach}, {Sheth}, \&
  {Teplitz}}]{draine2007}
{Draine}, B.~T., {Dale}, D.~A., {Bendo}, G., {et~al.} 2007, \apj, 663, 866

\bibitem[{{Fallscheer} {et~al.}(2009){Fallscheer}, {Beuther}, {Zhang}, {Keto},
  \& {Sridharan}}]{fallscheer2009}
{Fallscheer}, C., {Beuther}, H., {Zhang}, Q., {Keto}, E., \& {Sridharan}, T.~K.
  2009, \aap, 504, 127

\bibitem[{{Gaume} {et~al.}(1995){Gaume}, {Goss}, {Dickel}, {Wilson}, \&
  {Johnston}}]{gaume1995b}
{Gaume}, R.~A., {Goss}, W.~M., {Dickel}, H.~R., {Wilson}, T.~L., \& {Johnston},
  K.~J. 1995, \apj, 438, 776

\bibitem[{{Hildebrand}(1983)}]{hildebrand1983}
{Hildebrand}, R.~H. 1983, \qjras, 24, 267

\bibitem[{{Hoffman} {et~al.}(2003){Hoffman}, {Goss}, {Palmer}, \&
  {Richards}}]{hoffman2003}
{Hoffman}, I.~M., {Goss}, W.~M., {Palmer}, P., \& {Richards}, A.~M.~S. 2003,
  \apj, 598, 1061

\bibitem[{{Hofner} {et~al.}(2001){Hofner}, {Wiesemeyer}, \&
  {Henning}}]{hofner2001}
{Hofner}, P., {Wiesemeyer}, H., \& {Henning}, T. 2001, \apj, 549, 425

\bibitem[{{Ilee} {et~al.}(2013){Ilee}, {Wheelwright}, {Oudmaijer}, {de Wit},
  {Maud}, {Hoare}, {Lumsden}, {Moore}, {Urquhart}, \& {Mottram}}]{ilee2013}
{Ilee}, J.~D., {Wheelwright}, H.~E., {Oudmaijer}, R.~D., {et~al.} 2013, \mnras,
  429, 2960

\bibitem[{{Jenkins}(2004)}]{jenkins2004}
{Jenkins}, E.~B. 2004, in Origin and Evolution of the Elements, ed.
  A.~{McWilliam} \& M.~{Rauch}, 336

\bibitem[{{Keto}(1991)}]{keto1991}
{Keto}, E.~R. 1991, \apj, 371, 163

\bibitem[{{Knez} {et~al.}(2009){Knez}, {Lacy}, {Evans}, {van Dishoeck}, \&
  {Richter}}]{knez2009}
{Knez}, C., {Lacy}, J.~H., {Evans}, II, N.~J., {van Dishoeck}, E.~F., \&
  {Richter}, M.~J. 2009, \apj, 696, 471

\bibitem[{{Kraus} {et~al.}(2006){Kraus}, {Balega}, {Elitzur}, {Hofmann},
  {Preibisch}, {Rosen}, {Schertl}, {Weigelt}, \& {Young}}]{kraus2006}
{Kraus}, S., {Balega}, Y., {Elitzur}, M., {et~al.} 2006, \aap, 455, 521

\bibitem[{{Krumholz} {et~al.}(2007{\natexlab{a}}){Krumholz}, {Klein}, \&
  {McKee}}]{krumholz2007a}
{Krumholz}, M.~R., {Klein}, R.~I., \& {McKee}, C.~F. 2007{\natexlab{a}}, \apj,
  665, 478

\bibitem[{{Krumholz} {et~al.}(2007{\natexlab{b}}){Krumholz}, {Klein}, \&
  {McKee}}]{krumholz2006b}
{Krumholz}, M.~R., {Klein}, R.~I., \& {McKee}, C.~F. 2007{\natexlab{b}}, \apj,
  656, 959

\bibitem[{{Krumholz} {et~al.}(2009){Krumholz}, {Klein}, {McKee}, {Offner}, \&
  {Cunningham}}]{krumholz2009}
{Krumholz}, M.~R., {Klein}, R.~I., {McKee}, C.~F., {Offner}, S.~S.~R., \&
  {Cunningham}, A.~J. 2009, Science, 323, 754

\bibitem[{{Kuiper} {et~al.}(2011){Kuiper}, {Klahr}, {Beuther}, \&
  {Henning}}]{kuiper2011}
{Kuiper}, R., {Klahr}, H., {Beuther}, H., \& {Henning}, T. 2011, \apj, 732, 20

\bibitem[{{Kuiper} {et~al.}(2012){Kuiper}, {Klahr}, {Beuther}, \&
  {Henning}}]{kuiper2012}
{Kuiper}, R., {Klahr}, H., {Beuther}, H., \& {Henning}, T. 2012, \aap, 537,
  A122

\bibitem[{{Linz} {et~al.}(2009){Linz}, {Henning}, {Feldt}, {Pascucci}, {van
  Boekel}, {Men'shchikov}, {Stecklum}, {Chesneau}, {Ratzka}, {Quanz},
  {Leinert}, {Waters}, \& {Zinnecker}}]{linz2009}
{Linz}, H., {Henning}, T., {Feldt}, M., {et~al.} 2009, \aap, 505, 655

\bibitem[{{Linz} {et~al.}(2005){Linz}, {Stecklum}, {Henning}, {Hofner}, \&
  {Brandl}}]{linz2005}
{Linz}, H., {Stecklum}, B., {Henning}, T., {Hofner}, P., \& {Brandl}, B. 2005,
  \aap, 429, 903

\bibitem[{{Mathis} {et~al.}(1977){Mathis}, {Rumpl}, \&
  {Nordsieck}}]{mathis1977}
{Mathis}, J.~S., {Rumpl}, W., \& {Nordsieck}, K.~H. 1977, \apj, 217, 425

\bibitem[{{Maury} {et~al.}(2010){Maury}, {Andr{\'e}}, {Hennebelle}, {Motte},
  {Stamatellos}, {Bate}, {Belloche}, {Duch{\^e}ne}, \& {Whitworth}}]{maury2010}
{Maury}, A.~J., {Andr{\'e}}, P., {Hennebelle}, P., {et~al.} 2010, \aap, 512,
  A40

\bibitem[{{McKee} \& {Ostriker}(2007)}]{mckee2007}
{McKee}, C.~F. \& {Ostriker}, E.~C. 2007, \araa, 45, 565

\bibitem[{{McKee} \& {Tan}(2003)}]{mckee2003}
{McKee}, C.~F. \& {Tan}, J.~C. 2003, \apj, 585, 850

\bibitem[{{Minier} {et~al.}(2000){Minier}, {Booth}, \& {Conway}}]{minier2000}
{Minier}, V., {Booth}, R.~S., \& {Conway}, J.~E. 2000, \aap, 362, 1093

\bibitem[{{Minier} {et~al.}(2001){Minier}, {Conway}, \& {Booth}}]{minier2001}
{Minier}, V., {Conway}, J.~E., \& {Booth}, R.~S. 2001, \aap, 369, 278

\bibitem[{{Moscadelli} {et~al.}(2009){Moscadelli}, {Reid}, {Menten},
  {Brunthaler}, {Zheng}, \& {Xu}}]{moscadelli2009}
{Moscadelli}, L., {Reid}, M.~J., {Menten}, K.~M., {et~al.} 2009, \apj, 693, 406

\bibitem[{{Ossenkopf} \& {Henning}(1994)}]{ossenkopf1994}
{Ossenkopf}, V. \& {Henning}, T. 1994, \aap, 291, 943

\bibitem[{{Pestalozzi} {et~al.}(2009){Pestalozzi}, {Elitzur}, \&
  {Conway}}]{pestalozzi2009}
{Pestalozzi}, M.~R., {Elitzur}, M., \& {Conway}, J.~E. 2009, \aap, 501, 999

\bibitem[{{Pestalozzi} {et~al.}(2004){Pestalozzi}, {Elitzur}, {Conway}, \&
  {Booth}}]{pestalozzi2004}
{Pestalozzi}, M.~R., {Elitzur}, M., {Conway}, J.~E., \& {Booth}, R.~S. 2004,
  \apjl, 603, L113

\bibitem[{{Peters} {et~al.}(2010){Peters}, {Klessen}, {Mac Low}, \&
  {Banerjee}}]{peters2010b}
{Peters}, T., {Klessen}, R.~S., {Mac Low}, M.-M., \& {Banerjee}, R. 2010, \apj,
  725, 134

\bibitem[{{Puga} {et~al.}(2010){Puga}, {Mar{\'{\i}}n-Franch}, {Najarro},
  {Lenorzer}, {Herrero}, {Acosta Pulido}, {Chavarr{\'{\i}}a}, {Bik}, {Figer},
  \& {Ram{\'{\i}}rez Alegr{\'{\i}}a}}]{puga2010}
{Puga}, E., {Mar{\'{\i}}n-Franch}, A., {Najarro}, F., {et~al.} 2010, \aap, 517,
  A2

\bibitem[{{Qiu} {et~al.}(2011){Qiu}, {Zhang}, \& {Menten}}]{qiu2011}
{Qiu}, K., {Zhang}, Q., \& {Menten}, K.~M. 2011, \apj, 728, 6

\bibitem[{{Reid} \& {Wilson}(2005)}]{reid2005}
{Reid}, M.~A. \& {Wilson}, C.~D. 2005, \apj, 625, 891

\bibitem[{{Sandell} {et~al.}(2009){Sandell}, {Goss}, {Wright}, \&
  {Corder}}]{sandell2009}
{Sandell}, G., {Goss}, W.~M., {Wright}, M., \& {Corder}, S. 2009, \apjl, 699,
  L31

\bibitem[{{Sandell} \& {Sievers}(2004)}]{sandell2004}
{Sandell}, G. \& {Sievers}, A. 2004, \apj, 600, 269

\bibitem[{{Sandell} \& {Wright}(2010)}]{sandell2010}
{Sandell}, G. \& {Wright}, M. 2010, \apj, 715, 919

\bibitem[{{Schreyer} {et~al.}(2002){Schreyer}, {Henning}, {van der Tak},
  {Boonman}, \& {van Dishoeck}}]{schreyer2002}
{Schreyer}, K., {Henning}, T., {van der Tak}, F.~F.~S., {Boonman}, A.~M.~S., \&
  {van Dishoeck}, E.~F. 2002, \aap, 394, 561

\bibitem[{{Simon} {et~al.}(2000){Simon}, {Dutrey}, \& {Guilloteau}}]{simon2000}
{Simon}, M., {Dutrey}, A., \& {Guilloteau}, S. 2000, \apj, 545, 1034

\bibitem[{{Stahler} \& {Palla}(2005)}]{stahler2005}
{Stahler}, S.~W. \& {Palla}, F. 2005, {The Formation of Stars} (ISBN
  3-527-40559-3.~Wiley-VCH)

\bibitem[{{Thi} \& {Bik}(2005)}]{thi2005}
{Thi}, W.-F. \& {Bik}, A. 2005, \aap, 438, 557

\bibitem[{{Vaidya} {et~al.}(2009){Vaidya}, {Fendt}, \& {Beuther}}]{vaidya2009}
{Vaidya}, B., {Fendt}, C., \& {Beuther}, H. 2009, \apj, 702, 567

\bibitem[{{van der Tak} {et~al.}(2000){van der Tak}, {van Dishoeck}, \&
  {Caselli}}]{vandertak2000}
{van der Tak}, F.~F.~S., {van Dishoeck}, E.~F., \& {Caselli}, P. 2000, \aap,
  361, 327

\bibitem[{{Wheelwright} {et~al.}(2010){Wheelwright}, {Oudmaijer}, {de Wit},
  {Hoare}, {Lumsden}, \& {Urquhart}}]{wheelwright2010}
{Wheelwright}, H.~E., {Oudmaijer}, R.~D., {de Wit}, W.~J., {et~al.} 2010,
  \mnras, 408, 1840

\bibitem[{{Willner}(1976)}]{willner1976}
{Willner}, S.~P. 1976, \apj, 206, 728

\bibitem[{{Wolfire} \& {Cassinelli}(1987)}]{wolfire1987}
{Wolfire}, M.~G. \& {Cassinelli}, J.~P. 1987, \apj, 319, 850

\bibitem[{{Zhang} {et~al.}(2005){Zhang}, {Hunter}, {Brand}, {Sridharan},
  {Cesaroni}, {Molinari}, {Wang}, \& {Kramer}}]{zhang2005}
{Zhang}, Q., {Hunter}, T.~R., {Brand}, J., {et~al.} 2005, \apj, 625, 864

\bibitem[{{Zinnecker} \& {Yorke}(2007)}]{zinnecker2007}
{Zinnecker}, H. \& {Yorke}, H.~W. 2007, \araa, 45, 481

\end{thebibliography}

\end{document}